\documentclass[12pt]{iopart}
\eqnobysec
\usepackage{graphicx}
\usepackage{color}

\newcommand{\be}{\begin{equation}}
\newcommand{\ee}{\end{equation}}

\newcommand{\vecp}{{\mathbf p}}
\newcommand{\vecq}{{\mathbf q}}

\newcommand{\x}{{\mathbf x}}

\newcommand{\y}{{\mathbf y}}

\newcommand{\m}{{\mathbf m}}

\newcommand{\J}{{\mathbf J}}

\newcommand{\B}{\mathbf{B}}

\newcommand{\mH}{{\mathbf H}}

\newcommand{\M}{{\mathbf M}}

\newcommand{\Id}{{\mathbf I}}

\newcommand{\der}{\partial}

\newcommand{\vct}[1]{\ensuremath\mbox{\boldmath$ #1 $}}
\newcommand{\Vxi}{{\vct \xi}}

\newcommand{\Valpha}{{\vct \alpha}}

\begin{document}

\title{Semiclassical energy transition  of driven chaotic systems:
phase coherence on scar disks}

\author {Alfredo M. Ozorio de Almeida\footnote{ozorio@cbpf.br}}
\address{Centro Brasileiro de Pesquisas Fisicas,
Rua Xavier Sigaud 150, 22290-180, Rio de Janeiro, R.J., Brazil}

\begin{abstract}

A trajectory segment in an energy shell, which combines to form a closed curve with a segment in another canonically driven energy shell, 
adds an oscillatory semiclassical contribution to the smooth classical background 
of the quantum probability density for a transition between their energies. 
If either segment is part of a Bohr-quantized periodic orbit of either shell, the centre of its endpoints lies on a scar disk of the spectral Wigner function for single static energy shell and the contribution to the transition is reinforced by phase coherence. The exact representation of the transition density as an integral over spectral Wigner functions, which was previously derived for the special case where the system undergoes a reflection in phase space, is here generalized to arbitrary unitary transformations. If these are generated continuously by a driving Hamiltonian, there will be a finite lapse in the driving time for the transition to start, until the initially nested shells touch each other and then start to overlap.
 
The stationary phase evaluation of the multidimensional integral for the transition density selects the pair of matching trajectory segments on each shell, which close to form a piecewise smooth compound orbit. Each compound orbit shows up as a fixed point of a product of mappings that generalize Poincar\'e maps on the intersection of the shells. Thus, the closed compound orbits are isolated if the original Hamiltonian is chaotic. The actions of the compound orbits depend on the driving time, or on any other parameter of the transformation of the original eigenstates.

\end{abstract}

\maketitle

\section{Introduction}

Notwithstanding great progress in the experimental investigation and manipulation of quantum systems, it remains the rule that
initial states tend to be simple coherent states, or eigenstates of integrable Hamiltonians, such as those
of Rydberg atoms. There is no impediment, in principle, to evolve instead a highly excited eigenstate of a complex classically chaotic system 
and so to measure the probability of a given energy transition. Nonetheless, along with the experimental problem of preparing such a state,  
one faces the theoretical difficulty of coping with its, so far, only partially understood static features, let alone its evolution.
It is proposed here that a first step on the theoretical front is to develop the semiclassical (SC) framework of transitions between
coarsegrained microcanonical ensembles of eigenstates, in other words, averages over mixtures of states in energy ranges 
that are classically very narrow, while still comprising many eigenenergies. Such mixtures have provided important insight 
on the static properties of quantum chaotic systems and the aim here is to start to elucidate the underlying structure of trajectory segments
that determine semiclassically their reaction to external stimuli. 

The discovery of the scars of isolated periodic orbits in classically chaotic systems by Heller in 1984 \cite{Helscar} jostled the view that
representations of their eigenfunctions should only reflect uniform ergodic coverage of the corresponding energy shell, as, for instance, in the conjecture of Voros and Berry \cite{Vor76,Ber77a}. There is a certain analogy of the way these localized
classical structures affect the semiclassical (SC) static properties of eigenstates of chaotic systems to the more complex interaction of energy shells and trajectories that are here shown to underlie energy transitions of these eigenstates, when driven by the action of an external Hamiltonian, or directly by an unitary operator. Again, the broad classical picture of the $E\rightarrow E'$ transition depends exclusively on the smooth $E$-shell and the driven $E'$-shell, as studied in \cite{transE} (henceforth referred to as {\bf I}), but further SC refinement is shown to depend on a discrete set of isolated closed {\it compound orbits}, combining trajectory segments on both shells. 

The phase reinforcement of the contribution of a periodic orbit, which obeys an appropriate variant of Bohr quantization, to a mixed state in a narrow energy window, was derived semiclasically (SC) by Bogomolny \cite{Bogscars}. Then Berry \cite{Ber89} transported these results into phase space, where scars became a feature of the spectral Wigner function (a local energy average over pure state Wigner functions \cite{Wigner}) of classically chaotic systems. Later it was shown \cite{Report} that caustics in the interior of the energy shell touch it along each periodic orbit, so there is a profusion of caustics separating the multiple contributions  to the ordinary (complex exponential) SC spectral Wigner function.  
No previous attention appears to have been given to the two-dimensional (2-D) {\it scar disk}, which resembles a Wigner function for a single degree of freedon and is reinforced by phase coherence together with the periodic orbit that bounds it. It will be shown that selected energy parameters and unitary transformations, at which a scar disk is sampled, may be notable for increased  energy transitions. 

The action of an external quantum Hamiltonian drives the eigenstates through a continuous group of unitary transformations and the spectral Wigner function along with them, while the corresponding external classical Hamiltonian drives canonically the energy shells and trajectory segments, which are the ingredients of the SC approximation of the driven spectral Wigner function. The eigenergies are not altered in the driven Hamiltonian and likewise the driven energy shell keeps the same energy constant for the driven classical Hamiltonian, but the original Hamiltonian is not usually constant along a driven trajectory segment, nor is the driven Hamiltonian constant along a trajectory segment of the original Hamiltonian. 
Since all the points in an original $E'$-shell are transported to the driven $E'$-shell, the only contribution to the $E\mapsto E'$ classical transition is due to trajectory segments with endpoints on the intersection of the original $E$-shell with the driven $E'$-shell. Thus, the $E'$-shell acts as a generalized Poincar\'e section of the $E$-shell.  

The special role played in the Weyl-Wigner representation by the continuous set of unitary operators that correspond to classical reflections about any point in phase space, i.e. dislocated parity operators \cite{Grossmann,Royer}, was instrumental for an initial investigation of the probability of $E\mapsto E'$ energy transitions of quantum systems that are driven by these transformations. No distinction between regular, mixed or hard chaotic motion arose in the {\it classical approximation} in {\bf I}, constructed solely on the geometry of the classical $E$-shell, that of the classical reflection of the $E'$-shell and their intersection. 
Semiclassically, this may be considered as the contribution of very short trajectory segments, to which are here added the fully oscillatory SC contributions of long segments on both shells with endpoints lying on their intersection. 

It is important to note that the phase space reflections in {\bf I} were treated as sudden transformations and the same is true of their generalization here to general unitary operators,
even if generated by an external Hamiltonian. Indeed, the theory depends on the trajectories of the static initial $E$-shell 
and the final static $E'$-shell, once the driving Hamiltonian ceases to act. Therefore, it is assumed that the whole evolution is concluded
in a time scale that is much smaller than that of the intrinsic motion, as is usual in the logic gates of quantum information theory, 
or the the operations in nuclear magnetic resonance (NMR).

This paper presents the semiclassical role played by piecewise smooth {\it compound orbits} \cite{OzBro16}, formed by a trajectory segment in the original $E$-shell and joined at both tips to a segment in a driven $E'$-shell. Just like true periodic orbits in a single energy shell, these compound orbits will generically be isolated if the return maps to the intersection of the shells are chaotic. 
The compound orbits add oscillations to the smooth classical background presented in {\bf I}, 
as a parameter of the driving unitary transformation of the system is varied. 

This enriched picture results from the stationary evaluation of the exact integral for the energy transition probability density (or transition density for short) between a pair of coarsegrained energies. It was derived in {\bf I} from an exact integral identity for pure state Wigner functions, but here it is generalized to any unitary transformation with a classically evolving Weyl propagator. Beyond the previous reflections, this includes phase space translations (the Heisenberg group) and the metaplectic group, corresponding to symplectic (linear canonical) phase space transformations. Even for general unitary transformations, it is still valid semiclassically to evolve the Hamiltonian classically, with its trajectories on the evolved $E'$-shell, and thence to reconstruct the driven spectral Wigner function. The integral in which it enters for the transition density is then evaluated within the stationary phase approximation, which is again semiclassical. 

The general structure of trajectories, which contribute either to the energy transition density, 
or to the spectral Wigner function is quite similar in the special case of reflections and indeed the square modulus of the spectral Wigner function coincides with the diagonal density (describing permanence, rather than a transition). The crucial difference
to be kept in mind is that generally one integrates over Wigner functions, so that one is concerned with
dominant behaviour, such as at caustics. In contrast, an energy transition is generated by a definite unitary transformation and in no way
is it overshadowed by the transition specified by another parameter, be it an alternative time or another reflection centre. 
\footnote{It should be recalled that the direct point by point measurement of the Wigner function is achieved by repeating each specific
reflection of the eigenstate many times so as to evaluate its expectation \cite{Bertet02}.}

The caustics of the spectral Wigner function have lower dimension than the full $(2N)$-dimensional volume interior to an energy shell,
so there is a high probability that the stationary centre of a SC contributing compound avoids caustics in the interior 
of the appropriate energy shells. One can estimate energy transitions for continuous groups of unitary transformations starting from the identity, such as generated by any external driving Hamiltonian. There will be a finite interval before the driven $E'$-shell touches
the stationary $E$-shell, provoking a caustic peak in the time dependence of the transition density.

The exact expression for the energy transition density  to the driven by a general unitary transformation is derived in the following section as an integral over the spectral Wigner functions for the energy $E$ and the energy $E'$. The focus on scars limited previous derivations of the SC spectral Wigner function to the neighbourhood of periodic orbits\cite{Ber89,Report}, whereas here open trajectory segments are required. Thus, section 3 extends the SC approximation of spectral Wigner functions and their evolution, emphasizing the eventual phase reinforcement, if a trajectory segment has its centre lying on the 2-D scar disk bounded by a quantized periodic orbit. The multidimensional stationary phase approximation then provides the contribution of a compound orbit in section 4. Generally, there is a single contributing compound orbit for
each stationary centre, but multiple windings interfere positively if it lies in a scar disk, as discussed in section 5. Finally, this is completed with the full picture of general transition densities in section 6, including the classical term derived in {\bf I} together with the contributions of the compound orbits.

Over fifty years of collaboration with Michael Berry and the wonderful group that grew around him in Bristol has taught me a lot more than Wigner functions. I hope to continue this interaction way beyond the landmark of his eightieth birthday!

\section{Energy transitions driven by general unitary operators}

Consider the evolution of pure eigenstates $|k\rangle$ of a Hamiltonian $\hat H$ with energy $E_k$, under the action of a one parameter family of unitary operators
\be
\hat{U}(\tau) = \e^{-i\tau{\hat\Lambda}/\hbar}~,
\ee
where $\hat\Lambda$ is the driving Hamiltonian.
The probability of a transition to a state $|l\rangle$ in the driving time $\tau$ is
\begin{eqnarray}
P_{kl}(\tau) &= |\langle k|\hat{U}(\tau)|l\rangle|^2 = \langle k|\hat{U}(\tau)|l\rangle \langle l|\hat{U}^\dagger(\tau)|k\rangle 
\\ \nonumber  
&={\rm tr}~\hat{U}(\tau)|l\rangle\langle l|\hat{U}(\tau)^\dagger|k\rangle\langle k| = {\rm tr}~|l\rangle\langle l|(\tau)|k\rangle\langle k| ~,
\label{Ptrans0}
\end{eqnarray}
where the projector $|k\rangle\langle k|$ is a pure state density operator and one defines the driven pure state
\be
|l\rangle\langle l|(\tau) \equiv \hat{U}(\tau)|l\rangle\langle l|\hat{U}(\tau)^\dagger ~.
\ee

Let us introduce the Wigner-Weyl representation, within the corresponding classical phase space $\bf{R}^{2N}$ with its points 
$\{\x=(\vecp, \vecq)\}$, so that the pure states $|l\rangle\langle l|$ and $|l\rangle\langle l|(\tau)$ are represented  by the Wigner functions $W_k(\x)$ and $W_l(\x|\tau)$. Then the transition probability is obtained as
\be
P_{kl}(\tau) = (2\pi\hbar)^{N} \int {\rm d}^{2N}\x ~ W_k(\x)~W_l(\x|\tau) ~.
\label{ptrans1}
\ee
The propagation of the Wigner function generated by the Heisenberg operators
\be
{\hat T}_\Vxi = \exp\left[\frac{2i}{\hbar}\Vxi \wedge {\hat \x}\right] 
= \int {\rm d}^{N} \vecq ~|\vecq+ \Vxi_{\vecq}\rangle\langle \vecq- \Vxi_{\vecq}| \exp\left[\frac{2i}{\hbar}\Vxi_\vecp \cdot {\vecq}\right]  ~,
\ee
is purely classical. (Here $\Vxi$ is the halfchord for the corresponding phase space translation $\x \mapsto \x+ 2\Vxi$.)
\footnote{In fact, the translation operators are the basis for the chord representation, equivalent to the Wigner-Weyl representation, so one could construct the whole SC theory for energy transitions on the chord function instead of the Wigner function, but there is no immediate advantage, being that the latter are much more familiar. Here, we adopt the notation in \cite{SarOA19} in which the {\it halfchord} $\Vxi$ lables half the translation, instead of the full translation as in all my previous papers.}
This is also the case of the phase space reflection operator $\hat{U} =\hat{R}_{\x}$, that is,
\be
\fl \hat{R}_\x = \int \frac{{\rm d}^{2N} \Vxi}{(2\pi\hbar)^N} ~ {\hat T}_\Vxi  ~  \exp\left[\frac{2i}{\hbar}\x \wedge {\Vxi}\right] 
= \int {\rm d}^{N} \Vxi_{\vecq} ~|\vecq+ \Vxi_{\vecq}\rangle\langle \vecq- \Vxi_{\vecq}| ~
\exp\left[-\frac{2i}{\hbar}\Vxi_\vecp \cdot {\vecq}\right]  ~.
\ee  
In this case, it is recognized that \eref{ptrans1} is a generalization of the Wigner function identity on which {\bf I} was constructed. 
The list of operators, which classically evolve Wigner functions is completed by the metaplectic operators, which correspond to symplectic, i.e. linear canonical transformations in phase space. 

In general, the SC approximation of a driven eigenstate $W_l(\x|\tau)$ does not equal its simple classical evolution, which is known as the truncated Wigner approximation (TWA) \cite{Polk,Tit}, but it will be justified in the following section that it is identified
with the Wigner function of the $l$'th eigenstate of the driven Hamiltonian: 
\be
{\hat H}(\tau) \equiv \hat{U}(\tau){\hat H}~\hat{U}(\tau)^\dagger ~,
\label{evHam}
\ee
with the same energy $E_l$. In other words, the classical propagation of the Hamiltonian is SC superior to the direct classical
propagation of its eigenstates (because the latter may have highly oscillatory Wigner functions).

Now, one coarsegrains the transition probability as in {\bf I} by invoking the spectral Wigner function 
\cite{Ber89,Ber89b,Report}
\be
W_E(\x,\epsilon) \equiv (2\pi\hbar)^N \sum_k \delta_\epsilon(E-E_k)~ W_k(\x),
\label{spectral1}
\ee
which represents the spectral density operator
\be
{\hat \rho}_E(\epsilon) \equiv \sum_k \delta_\epsilon(E-E_k)~ |k\rangle \langle k|
\ee
for a classically narrow energy range $\epsilon$ centred on the energy E. Here 
\be
\delta_\epsilon(E)\equiv \frac{1}{\pi} ~ \frac{\epsilon}{\epsilon^2 + E^2}
\label{widelta}
\ee 
integrates as a Dirac $\delta$-function
and likewise
\be
W_E(\x,\epsilon|\tau) \equiv (2\pi\hbar)^N \sum_k \delta_\epsilon(E-E_k)~ W_k(\x|\tau)
\label{spectralev}
\ee
is the spectral Wigner function for the driven system. 
Then the probability density for the transition $E \mapsto E'$
\be
P_{EE'}(\tau, \epsilon) \equiv {\rm tr} ~{\hat \rho}_E(\epsilon) ~{\hat \rho}_{E'}(\epsilon|\tau)
\label{optrans}
\ee
results from the summation over both indices in \eref{ptrans1}  as
\begin{eqnarray}
 \fl P_{EE'}(\tau, \epsilon)  
= (2\pi\hbar)^{N}\sum_{k,l}  \delta_\epsilon(E-E_k)~\delta_\epsilon(E'-E_l) \int {\rm d}^{2N}\x ~ W_k(\x)~W_l(\x|\tau) \\ \nonumber
= \frac{1}{(2\pi\hbar)^{N}} \int {\rm d}^{2N}\x ~ W_E(\x, \epsilon)~W_{E'}(\x,\epsilon|\tau) ~.
\label{ptrans2}                                                                  
\end{eqnarray}

The limitation to the continuous group of unitary transformations generated by a particular external Hamiltonian is in no way essential.
Indeed, it should be noted that the reflection operator $\hat{R}_{\x}$ is never close to the identity $\hat \Id$, so that the reflections
do not form a group and it is only in combination with the Heisenberg group of translation operators ${\hat T}_\Vxi$ that the quantum version of the affine group is formed \cite{Coxeter,Report}. Thus, the translation operators will be the main example, for which the transition densities for general translations $2\Vxi$ take the specific form
\be
P_{EE'}(\Vxi, \epsilon) = \frac{1}{(2\pi\hbar)^{N}} \int {\rm d}^{2N}\x ~ W_E(\x, \epsilon)~W_{E'}(\x + 2\Vxi,\epsilon) ~,
\label{ptrans3}
\ee 
bearing in mind that the simplest driving Hamiltonian $\Lambda(\x) = {\Valpha}\wedge \x/2$ generates the subgroup of translations  
with the half chord $\Vxi(\tau)=\tau~{\Valpha}$. 

A further advantage of focusing  on the translations is that a reflection through a general point $\x$ combines a reflection through the origin with a translation by $2\Vxi=2\x$. If $H(\x)$ is symmetric with respect to the origin, there is no difference between the reflected energy shell and the shell merely translated by $2\x$. We can then keep in mind the structures of centre sections explored in {\bf I},
but in general the reflection centre $\x$ does not need to be close to the $E$-shell, where lies the caustic of the SC spectral Wigner function that was there analyzed. On the other hand, a subgroup of translations by $2\tau\Valpha$ of a nearby energy shell will take a short time interval until the smaller section touches the larger one from its inside. With further translations the section quickly grows into a large manifold, though it may still have the topology of a sphere, just as common Poincar\'e sections. One can expect most, if not all, periodic orbits to traverse the section, so that their successive intersections with the section become periodic points of the generalized Poincar\'e map. 

The transition density $P_{EE'}(\tau, \epsilon)$ is expressed in (2.13) as an integral over a spectral Wigner function and a driven spectral Wigner function. A first appraisal of the transition density is provided by the intersection of the pair of energy shells involved in the transition. It was shown in {\bf I} that near a caustic one can get away with an integral just over their intersection. For large sections one needs to add a full SC description of the spectral Wigner functions in terms of trajectory segments, each with a phase determined by its classical action.

\section{The spectral Wigner function, its scars and scar disks}

 The SC approximation for $W_E(\x,\epsilon)$ is based on trajectory segments in the $E$-shell with their tips
$\x_\pm$ centred on $\x$ \cite{Ber89,Ber89b,Report}. Let us return to its construction from the {\it Weyl propagator} $V(\x,t)$, which represents the intrinsic unitary evolution operator
\be
\hat{V}(t) = \e^{-it{\hat H}/ \hbar}~,
\label{intev}
\ee
that is,
\be
W_E(\x,\epsilon) = {\rm Re} \int_0^\infty \frac{{\rm d}t}{\pi\hbar}~\exp\left[ \frac{it}{\hbar}(E+i\epsilon)\right]~ V(\x,t) ~. 
\label{spectral2}
\ee  
In the simplest case, the Weyl propagator has the SC approximation
\begin{equation}
V(\x,t) \approx \frac{2^N}{|\det(\Id+ \M(\x,t))|^{1/2}}\>\>
\exp \left[ \frac{i}{\hbar}(S(\x,t))+ \hbar \pi \sigma)\right],
\label{Uweyl}
\end{equation}
if there is just a single classical trajectory 
centred on the point $\x$; otherwise there will be a superposition of similar terms.
The geometric part of the {\it centre} or {\it Weyl action} $S(\x,t)$ is just the symplectic area between
the trajectory $\tilde \x(t,\x_-)$ and the chord $2\Vxi=\x_+ -\x_-$,
as shown in Fig.1. From this, one subtracts $-Et$, where $E$ is the energy of the trajectory.

\begin{figure}
\centering
\includegraphics[width=.6\linewidth]{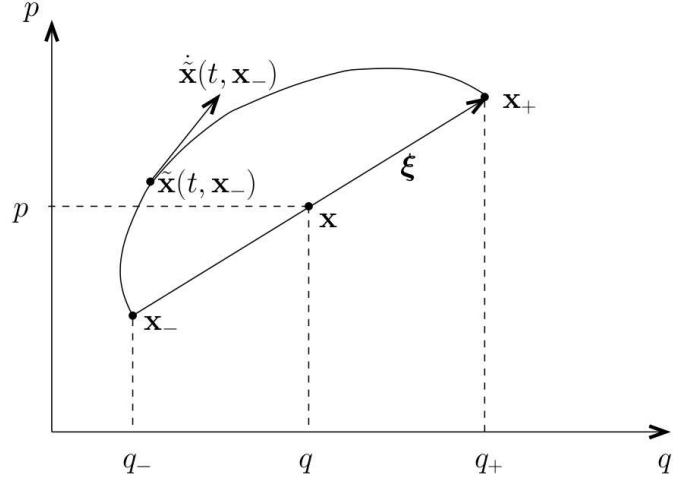}
\caption{The action, $S(\x,t)$, is just the symplectic area between
the trajectory from ${\x}^-$ to ${\x}^+$ and its chord $2\Vxi$, from which one subtracts $Et$,
where $E$ is the energy of the trajectory.}
\label{Fig1}
\end{figure}

The centre action specifies the classical canonical transformation, $\x_- \mapsto \x_+$,
corresponding to ${\hat V}(t)$ indirectly through \cite{Report}
\begin{equation}
2\Vxi = -\J \frac{\der S}{\der \x}, \>\>~~ \x_{\pm} = \x \pm \Vxi,
\label{centran}
\end{equation}
where the standard symplectic matrix is
\be
\J = \left(
\begin{array}{cc}
     0 & -\Id_N \\
     \Id_N & 0 
\end{array}
\right) 
\ee 
in terms of $N$-D blocks in the $\mathbf{p}$-space and the $\mathbf{q}$-space.
The linear approximation of this transformation near the $\x$-centred trajectory is defined
by the, {\it stability matrix}, or the {\it symplectic monodromy matrix} $\M(\x,t)$. This has the Cayley parametrization \cite{Report}:
\begin{equation}
\M(\x,t) = [\Id + \J\mathbf{B}(\x,t)]^{-1} [\Id - \J\mathbf{B}(\x,t)],
\label{Cayley}
\end{equation}
in terms of the symmetric Hessian matrix of the action
\be
\mathbf{B}(\x,t) \equiv \frac{1}{2}~ \frac{\der^2 S(\x,t)}{\der \x^2}
\label{Hessian}
\ee
and the full identity matrix $\Id$.
 
There may be multiple solutions to the variational problem that identifies trajectories with a given centre $\x$, 
so the actions may have many branches and these branches meet along  caustics where the semiclassical amplitude diverges. 
For short times there are no caustics \cite{Report} and $\sigma=0$.

The stationary phase evaluation of \eref{spectral2} singles out the trajectory segments on the $E$-shell with their tips $\x_\pm$ centred on $\x$.
Their times $t_j$ may be positive or negative , i.e. one also includes ${\tilde \x}(-t)$, exchanging the endpoints $\x_\pm$ centred on $\x$, because of taking the real part, 
with the result
\be
\fl W_E(\x,\epsilon) \approx \sqrt{\frac{2}{\pi\hbar}} \sum_j \frac{2^N}{[{\rm d}E/{\rm d}|t_j|~|\det(\Id+ \M(\x,t_j))|]^{1/2}} 
~ \e^{-\epsilon |t_j|/\hbar} ~\cos\left[ \frac{S_j(\x,E)}{\hbar} + \lambda_j \right] ~.
\label{spectral3}
\ee
The only change in the energy action branches $S_j(\x, E)$ with respect to the time action $S(\x,t)$, for the same trajectory segment
that is centred on $\x$, is the absence of the term $-Et$.
Thus the relation \eref{centran} between the centre $\x$ and the endpoints still holds for the branches of the energy action, even though
it is only the time action that plays the role of a generating function for the respective canonical transformation. It should be kept in mind
that the traversal time $t_j$ is also a function of $E$, though this is not shown explicitly to simplify the notation. 

Previous treatments of the spectral Wigner function have been mostly concerned with their scars in the neighbourhood of periodic orbits, so that the endpoints nearly coincide. Then it is possible to adopt time-energy variables locally, which limits the stability matrix to the remaining transverse variables. Even though distant endpoints cannot be so simply treated, a variant of the {\it velocity plane} introduced in {\bf I} is of help. Indeed, the 2-D plane spanned by the endpoint velocities $\dot{\x}_+$ and $\dot{\x}_-$ may be chosen for the $(p,q)$ coordinates of the $N$'th conjugate plane in a symplectic coordinate system for the $2N$-D phase space. Each of these velocities has a $(2N-1)$-D skew-orthogonal plane, \footnote{that is, the plane orthogonal to $\J\dot{\x}_\pm$,} identified with the pair of tangent planes to the energy shell at $\x_+$ and $\x_-$. The $2(N-1)$-D intersection of these planes is skew-orthogonal to the velocity plane, which it includes. 
Then an arbitrary symplectic coordinate system can be chosen, the $\y$ space, together with $(p,q)$, to complete the $2N$ symplectic coordinates for the full phase space \cite{Arnold}.

Following {\bf I} and originally \cite{AlmHan82} one introduces the pair of displaced Hamiltonians
\be
H_+ = H(\x+\Vxi) ~~~ {\rm and} ~~~ H_- =  H(\x-\Vxi) ~,
\label{section}
\ee
which can be considered as either functions of the centre $\x$, for a fixed halfchord $\Vxi$, or vice versa.
Then Hamilton's equations for the pair of basis vectors $\dot{\x}_\pm$ can be expressed as
\be
\dot{p}_\pm = - \frac{\der H_\pm}{\der q}  ~~~~ {\rm and}  ~~~~ \dot{q}_\pm =\frac{\der H_\pm}{\der p} ~,
\ee
whereas 
\be
\dot{\y}_\pm= \J\frac{\der H_\pm}{\der \y} = \J\frac{\der H}{\der \y}(\x_\pm) =0 ~.
\ee 
Furthermore, the Jacobian matrix for adopting the displaced Hamiltonians themselves as (nonsymplectic) coordinates on the velocity plane, 
instead of $(p,q)$, is just
\be
\frac{\der(H_+, H_-)}{\der(p,q)} =
\frac{\der H_+}{\der p}~ \frac{\der H_-}{\der q} - \frac{\der H_+}{\der q} ~ \frac{\der H_-}{\der p}
= \dot{\x}_+ \wedge \dot{\x}_-
\equiv \{H_+,H_-\} ~,
\label{PB}
\ee
which is identified as the full Poisson bracket of the pair of displaced Hamiltonians. 

Let us now consider the stability matrix for the linearized motion surrounding the chosen trajectory segment centred on $\x$ in the time $t_j$:
$\delta\x_+=\M(\x,t_j)~\delta\x_-$. This symplectic transformation takes a tangent plane to the $E$-shell at $\x_-$ to its tangent plane
at $\x_+$. If $\delta \y_- = 0$, that is, the only component of $\delta \x_-$ in the initial tangent plane lies along $\dot{\x}_-$, then the component of $\delta \x_+$ in the tangent plane at $\x_+$ follows $\dot{\x}_+$, so that we also have $\delta \y_+ = 0$. 
Conversely, all displacements on the initial tangent plane of the form $(\delta \y_-,0,0)$ (with no component on the velocity plane)
are transported to the final tangent plane and hence are expressed as $(\delta \y_+,0,0)$, since, 
again here there is no component on the velocity plane.
Thus,  $\delta\y_+ =\M_\y(\x,t_j)~\delta \y_-$. 
In short, if any displacement in the velocity plane propagates independently of all displacements
in the transverse $\y$-space, the stability matrix factors into a transverse block $\M_\y$ and a 2-D block
$\m$ in the velocity plane, which is identified with the full matrix $\M$ itself in the case of a single degree of freedom. 
In other words, $\M= \m \otimes \M_\y$, just as $\Id = \Id_2 \otimes \Id_\y$ and $\J = \J_2 \otimes \J_\y$, so that 
$\det(\Id+ \M(\x,t_j))=\det(\Id_2+ \m(\x,t_j))~\det(\Id_\y+ \M_\y(\x,t_j))$ in the denominator of \eref{spectral3}, noting that
for a $2\times 2$ symplectic matrix
\be
(\Id_2+ \m)^{-1} = [\det(\Id_2+ \m)]^{-1}~ (\Id_2+ \m^{-1}).
\label{invmat}
\ee

The energy derivative in the SC spectral Wigner function presuposes trajectories with a fixed centre $\x$, 
but it is more natural to fix the initial point $\x_-$ and to limit $\delta\x_-$ to the velocity plane, so that for $\x(t,\x_-)$ we impose
\be
\fl \delta\x = \frac{\der \x}{\der \x_-}~ \delta\x_- + \frac{\der \x}{\der t} ~ \delta t
= \frac{1}{2} ~(\Id_2+ \m(t_j))~  \delta \x_- + \frac{1}{2} ~ \dot{\x}_+(\x_-,t) ~\delta t = 0,
\ee
that is,
\be
\frac{\der \x_-}{\der t}(\x,t) = - (\Id_2+ \m)^{-1} ~ \dot{\x}_+(\x_-,t)
\ee
and hence, for fixed $\x$,
\begin{eqnarray}
\frac{{\rm d}E}{{\rm d}t}& = \frac{\der H}{\der \x_-} \cdot \frac{\der \x_-}{\der t}(\x,t)
=-(\J_2~\dot{\x}_-) \cdot (\Id_2+ \m)^{-1} ~ \dot{\x}_+  \\  \nonumber
&= [\det(\Id_2+ \m)]^{-1} ~ \dot{\x}_- \wedge (\dot{\x}_+ + \dot{\x}_-)  \\ \nonumber
&= [\det(\Id_2+ \m)]^{-1} ~ \{H_-, H_+\}, 
\end{eqnarray}
using \eref{invmat}. 

In conclusion, the SC approximation for the spectral Wigner function becomes
\be
\fl W_E(\x,\epsilon) \approx \sqrt{\frac{2}{\pi\hbar}} \sum_j \frac{2^N}{[~|\{H_-, H_+\}_j\det(\Id_\y+ \M_\y(\x,t_j))|]^{1/2}}  
~ \e^{-\epsilon |t_j|/\hbar}~\cos\left[ \frac{S_j(\x,E)}{\hbar} + \lambda_j \right] .
\label{spectral4}
\ee
In the case of a single freedom, there are no transverse $\y$-coordinates and hence no determinant remains in the denominator,
which becomes singular at the caustic as $\x$ touches the energy shell. The remaining Poisson bracket of the displaced Hamiltonians is proportional to the Poisson bracket of the pair of action variables in the SC expression for a pure state Wigner function in \cite{AlmHan82}. 

Averaging over oscillations in $|W_E(\x,\epsilon)|^2$, one obtains a classical envelope of the sum over all trajectory segments-$j$, 
in terms of the full denominators $\{H_+, H_+\}_j$. 
This is here a discrete sum, instead of the integral in {\bf I} (effectively over a continuum of $t=0$ trajectories), being that the transverse factor $|\det(\Id_\y+ \M_\y(\x,t_j))|$ in the denominator brings in the stability of each finite trajectory segment. 
Defining the {\it centre section} \cite{Report} (also reviewed in {\bf I}) of the $E$-shell by its reflection through the point $\x$, 
each contributing segment determines a fixed point of some iteration of a generalized Poincar\'e map on this section.
The exponential damping of the contributions to the spectral Wigner function with increasing period results in the general dominance of the
shortest segment. In the limit as the centre $\x$ approaches the energy shell, one must resolve the singularity of \eref{spectral3} of this
very short segment contribution, as reviewed in {\bf I}. Otherwise, there will be decreasing contributions, dampened by the exponent 
$\epsilon \>t_j(E)/\hbar$, such that the lable $j$ describes the successive intersections of the trajectory segment with the centre map.

Up to here, no special role has been played by periodic orbits, but it is clear for any centre $\x$, such that both endpoints $\x_\pm$ 
lie on a periodic orbit (o) with period  $\tau_{\rm o}$ and action $S_{\rm{o}}(E)$, that 
\be
S_j(\x, E)+ j'~ S_{\rm{o}}(E) = S_{j''}(\x,E) ~.
\ee  
Thus, the variant of Bohr quantization,
\be 
\frac{S_j(\x,E)}{\hbar} + \lambda_j =2k\pi ~,
\label{Bquant}
\ee
guarantees phase coherence among all repetitions for each centre of a periodic orbit chord. 

Of course if $N=1$ , this holds for all points 
within the $E$-shell (itself a periodic orbit) and, diminishing $\epsilon$, one converges onto the SC Wigner function for the quantized state.
The periodic orbit is a caustic and there are generally other internal caustics, which limit regions where the centres have different numbers of chords \cite{Ber77}. The special feature of the external caustic is its constant phase, so that it dominates any integral
over the Wigner function, whereas the internal caustics are usually washed away. 

For $N>1$, the locus of centres is again a 2-D disk, bounded by the periodic orbit orbit, but this may have a complex topology, for instance, a periodic orbit neighbouring a homoclinic orbit. Here, we need not concern ourselves with such complications, since long periods are dampened away 
by the finite energy width $\epsilon$, leaving only relatively few fairly simple closed curves. It is important to note
that the quantized phase reinforcement \eref{Bquant} holds for all points in the centre disk of a quantized periodic orbit. 
So far, attention for this {\it quantized scar disk} has been overshadowed by its caustic along the periodic orbit itself, 
which is singled out as the {\it scar}. It dominates integrals over the spectral Wigner function (just as in the case where $N=1$). 

The resulting picture from \cite{Ber89} further refined in \cite{Report} is that a pair of $(2N-1)$-D (fold) caustic surfaces touch each other tangentially along each periodic orbit. The first is the $E$-shell itself, whereas the other caustic limits the 'lateral' range of the centres of chords that might be considered as perturbations of those on the periodic orbit, that is, those centres for which the centre section is traversed 
by the given periodic orbit. According to \cite{Report} the width of the second caustic shrinks parabollicaly as the the periodic orbit is approached, but no effort has yet been made to follow this second caustic within the $E$-shell. In any case, the caustic certainly includes the 
2-D scar disk, which is the exclusive locus of quantized centres.

So far the focus has been on the original spectral Wigner function, but we also need its evolved form \eref{spectralev}. This is determined by a purely classical evolution of its argument in the simple cases of reflections and translations, which may be combined with symplectic transformations, but not general canonical evolutions generated by a driving Hamiltonian $\Lambda(\x)$ with higher than second order terms. Nonetheless, the SC approximation of the Weyl representation of the driven quantum Hamiltonian \eref{evHam} is merely 
$H(\x|\tau) = H(\tilde{\x}(-\tau,\x))$, that is, the driven classical Hamiltonian. On the other hand, the intrinsic evolution operator 
\eref{intev} generated by the driven Hamiltonian 
is simply
\be
\hat{V}(t|\tau) \equiv \hat{U}(\tau) ~\e^{-it{\hat H}/ \hbar}~\hat{U}(\tau)^\dagger = \exp\left[-\frac{it}{\hbar}\hat H(\tau)\right]~.
\ee
It follows that the corresponding SC approximation for the evolved Weyl propagator $V(\x,t|\tau)$ is just \eref{Uweyl} based on the classical trajectories of the driven classical Hamiltonian and consequently \eref{spectral2} supplies the SC evolved spectral Wigner function in the form as \eref{spectral3}.
\footnote{ In this way, the usual procedure of constructing SC approximations based on evolving Lagrangian manifolds is extended to structures of different dimensionality in phase space.}

Explicitly then the SC approximation for the evolved spectral Wigner function for the coarsegrained evolved $E'$-shell, $H(\x|\tau)=E'$, is
 \begin{eqnarray}
\fl W_{E'}(\x,\epsilon|\tau) \approx \sqrt{\frac{2}{\pi\hbar}} ~ \sum_j ~ 
& \frac{2^N}{[~|\{H_-, H_+\}_j(\tau) ~ \det(\Id_\y+ \M_{\y}(t_j|\tau))|]^{1/2}} ~  \\    \nonumber
&\exp\left[-\frac{\epsilon |t_j|}{\hbar}\right] ~  \cos\left[ \frac{S_j(\x,E'|\tau)}{\hbar} + \lambda_j \right] .
\label{spectral5}
\end{eqnarray}
Here the driven pair of displaced classical Hamiltonians $H_\pm(\x|\tau)=H(\x_\pm(t_j|\tau))$ has the Poisson bracket $\{H_+, H_+\}_j(\tau)$, 
though the dependence on the centre $\x$, as that of the stability matrix, is not shown explicitly. Indeed, one should keep in mind that even the time $t_j$ of each trajectory segment is $\x$-dependent and that the index-$j$ runs over positive and negative times. The derivative of the driven action $S_j(\x,E'|\tau)$ with respect to $\x$ supplies the driven chord $\Vxi_{j'}(\x|\tau)$ and hence the endpoints, which lie on the driven energy shell according to \eref{centran}.

\section{Stationary phase evaluation of the energy transition density}

Inserting the SC approximations (3.17) and (3.21) of the original and the evolved spectral Wigner functions in the exact expression (2.13) for the transition density, results in a sum of integrals over complex exponentials labled by all the windings of the trajectory segments lying on either energy shell, such that their centre is the integration variable $\x$:
\begin{eqnarray}
\fl P_{EE'}(\tau, \epsilon)  
\approx {\rm Re}~ \left(\frac{2}{\pi\hbar}\right)^{N+1}\sum_{j,j'} ~ 
\int  {\rm d}\x ~ \exp \left[\frac{i}{\hbar}(S_j(\x,E) + S_{j'}(\x,E'|\tau) + i (\lambda_j + \lambda_{j'})  \right]\\  \nonumber 
\frac{ \exp[-\epsilon(|t_j| + |t_{j'}|)/\hbar]} 
{| \{H_-, H_+\}_j \{H_-, H_+\}_{j'}(\tau)	 ~ \det(\Id_\y+ \M_\y(t_j)) \det(\Id_\y+ \M_\y(t_{j'}|\tau))|^{1/2}} ~ .
\label{statint}
\end{eqnarray}
The stationary phase condition for the integral in each term then singles out the stationary points $\x =\x_{jj'}$, for which
\be  
\frac{d}{d\x}[S_j(\x,E) + S_{j'}(\x,E'|\tau)] = 0.
\ee
According to \eref{centran} this implies that the halfchords $\Vxi_{j}(\x_{jj'}) + \Vxi_{j'}(\x_{jj'}|\tau) = 0$
(allowing for the notation $\Vxi(\x_{jj'}) = \Vxi_{j}(\x_{jj'}) = -\Vxi_{j'}(\x_{jj'}|\tau)$),
so that the pairs of segments centred on the stationary points join continuously at both their endpoints
\footnote{It should be noted that the coincidence of the pair of endpoints does not carry over to the velocities
$\dot{\x}_\pm(\x_{jj'}, t_j)$ and $\dot{\x}_\pm(\x_{jj'}, t_{j'})$, leading to different Poisson brackets in the the denominator of the integrand, which is just their skew product, according to \eref{PB}. Thus, one should also distinguish the variables
$\y(t_j)$ and $\y(t_{j'})$ that are transverse to the respective velocity planes.}  
\be
\x_+(\x_{jj'}, t_j) = \x_-(\x_{jj'}, t_j') ~~~{\rm and} ~~~ \x_+(\x_{jj'}, t_j') = \x_-(\x_{jj'}, t_j) ~.
\ee
In other words, each stationary phase contribution to the transition density is ascribed to a single continuous piecewise smooth trajectory joining both energy shells, represented by the thick curves in Fig. 2. This is an example of a (closed) {\it compound orbit}
\footnote{Originally constructed from trajectory segments in SC Weyl propagators in \cite{OzBro16}.} 
and its total action is
\be
S_{jj'} = S_j(\x_{jj'},E) + S_{j'}(\x_{jj'},E'|\tau) ~.
\ee  

The full evaluation of (4.2) by stationary phase requires the variation of the sum of the chord areas of both segments as their common centre $\x$ moves out from $\x_{jj'}$. Then they no longer fit together and the tips of their chords $2\Vxi_{j}(\x)$ and $2\Vxi_{j'}(\x|\tau)$, both centred on $\x=\x_{jj'} + \delta \x$, can be joined in an elongated polygonal figure of eight. Illustrated in Fig. 2, it has zero symplectic area because of the central symmetry.  This continuous displacement of the centre from zero to $\delta \x$ brings with it a continuous family of trajectory segments on the $E$-shell. The displacement of the endpoints $\delta \x_\pm(E)$ (shown in red in the figure) connect opposite corners of the figure of eight at $\x_\pm(\x_{jj'} + \delta \x)$ to the endpoints $\x_\pm(\x_{jj'})$ of the stationary chord. By the Poincar\'e-Cartan theorem \cite{Arnold,livro}, the elongated circuit formed by this family of trajectory segments also has zero symplectic area. The analogous construction, of an elongated circuit on the border of trajectory segments on the driven $E'$-shell, is closed by another pair of (red) displacements $\delta \x_\pm(E')$,
also with tips at $\x_\pm(\x_{jj'})$. Adding and subtracting these three circuits with zero symplectic area to the original pair of actions, it is verified that all that is left in the overall variation of the sum of the actions is the pair of small triangles, here illustrated in red.

\begin{figure}
\centering
\includegraphics[width=.6\linewidth]{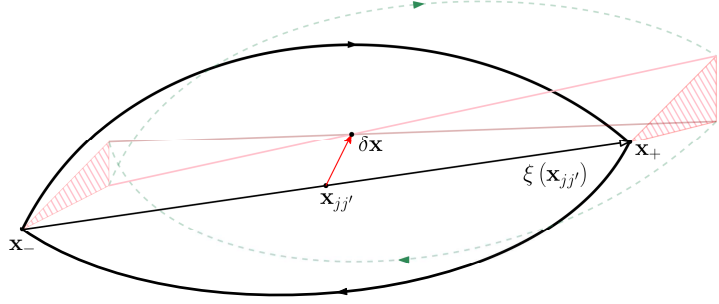}
\caption{The pair of segments centred on $\x_{jj'}$ meet to form the compound orbit, the piecewise continuous curve, 
but the ends of the pair of chords centred on $\x_{jj'} + \delta\x$ do not coincide. The variation with respect to the stationary action
$S_{jj'}$ is of second order in $\delta\x$ and equals the sum of the symplectic areas of the triangles (shown in red) at the tips of three chords.}
\label{Fig2}
\end{figure}
 
In terms of the Hessian matrices
\be
\B_j = \frac{1}{2}\frac{\der^2 S_j}{\der \x^2}(\x_{jj'},E) ~~~ {\rm and} ~~~ 
\B_{j'} = \frac{1}{2}\frac{\der^2 S_{j'}}{\der \x^2}(\x_{jj'},E'|\tau) ~,
\label{Hessian}
\ee
the variation of the endpoints is
\be
\delta \x_\pm( E) = (\Id \pm \J \B_j) \delta \x ~~~ {\rm and} ~~~ \delta \x_\pm( E') = (\Id \pm \J \B_{j'}) \delta \x. 
\ee
So the symplectic area $\Delta_1$ of the first triangle follows 
\begin{eqnarray}
2 \Delta_1 & = - \delta \x_-(E)\wedge \delta \x_+(E') = - [\J (\Id - \J \B_j) \delta \x]\cdot [(\Id + \J \B_{j'}) \delta \x]
\\   \nonumber
&= \delta \x \cdot (\J - \B_j - \B_{j'} - \B_j \J \B_{j'}) ~\delta \x
\end{eqnarray}
and adding an analogous expression for the second triangle supplies
\be
2(\Delta_1 + \Delta_2) = - \delta \x \cdot ( 2\B_j + 2\B_{j'} + \B_j \J \B_{j'} +  \B_{j'} \J \B_j)~ \delta \x~,
\ee
so that
\be
\frac{d^2}{d\x^2}[S_j(\x,E) + S_{j'}(\x,E'|\tau)] = 2\B_j + 2\B_{j'} + \B_j \J \B_{j'} +  \B_{j'} \J \B_j.
\label{Hessall}
\ee
Thus, defining the signature $\sigma_{jj'}$  of this matrix, the SC approximation for the contribution of the compound orbits to the transition density becomes 
\begin{eqnarray}
\fl P_{EE'}(\tau, \epsilon)  
\approx {\rm Re}~\left(\frac{2}{\pi\hbar}\right)^{N+1}\sum_{j,j'} ~  
\exp \left[\frac{i}{\hbar}S_{jj'} + i (\lambda_j + \lambda_{j'}+ \frac{\pi}{4}\sigma_{jj'}) -\frac{\epsilon}{\hbar}(|t_j| + |t_{j'}|)  \right]
 \\   \nonumber
\{| \{H_-, H_+\}_j \{H_-, H_+\}_{j'}(\tau)	 ~ \det(\Id_\y+ \M_\y(t_j)) \det(\Id_\y+ \M_\y(t_{j'}|\tau)) \\   \nonumber
~ \det(2\B_j + 2\B_{j'} + \B_j \J \B_{j'} +  \B_{j'} \J \B_j )|\}^{-1/2} ~ .
\label{SCtrpr}
\end{eqnarray}

It should be noted that the phases of the SC contributions of compound orbits preserve no specific imprint of the Wigner-Weyl representation. Indeed, the action $S_{jj'}$ of each closed compound orbit need not be decomposed into two centre actions
and one would obtain the same result in another representation. For instance, an inverse  Weyl-Wigner transformation of the spectral Wigner function furnishes the spectral density operator ${\hat \rho}_E(\epsilon)$ as $\langle \vecq_-|{\hat \rho}_E(\epsilon)|\vecq_+\rangle$,
its position representation. Again the SC approximation is based on trajectory segments, but constrained by their fixed end positions $\vecq_\pm$,
while the derivative of the actions supplies the end momenta $\vecp_\pm = \vecp(\vecq_\pm,E)$. 
Then the trace of the product operator in \eref{optrans} becomes an integral over $\vecq_-$ and $\vecq_+$ instead of $\x$ 
and the stationary condition is just
\be
\vecp(\vecq_+,E) = \vecp(\vecq_-,E'|\tau) ~~~ {\rm and} ~~~ \vecp(\vecq_+,E'|\tau) = \vecp(\vecq_-,E) ~.
\ee
Hence, one obtains again contributions from closed compound orbits with the same action as before. On the other hand,
the amplitudes of compound orbits will not be immediately translatable into that in (4.10) and it may be worthwhile
to try to rework these alternatives into a simpler more compact form in the future.

\section{Reinforcement on a scar disk}

The finite width of the coarsegrained energy shells dampens exponentially the compound orbits for which 
$|t_j| + |t_{j'}| > \hbar/\epsilon$, effectively cutting off their contribution to the SC sum for the transition density.
Thus, generally, the transition density is dominated by the contribution of the smallest compound orbits, 
unless there is some positive phase interference.
No reinforcement through phase coherence of contributions of Bohr-quantization of the compound orbits themselves is available, 
since the segments are uniquely traversed for each $j$ and $j'$. 
Nonetheless, phase reinforcement occurs if the stationary point $\x_{jj'}$ lies on a scar disk 
of a periodic orbit, that is, either the $j$-segment is an arc of a Bohr-quantized periodic orbit in the $E$-shell, or the $j'$-segment is an arc 
of a Bohr-quantized periodic orbit in the evolved $E'$-shell. Then successive returns of the quantized periodic orbit are again compound orbits 
centred on $\x_{jj'}$, with the same chord $\Vxi(\x_{jj'})$ and action differences that alter the phase by multiples of $2\pi$, just as for
the scar disk spectral Wigner function itself \eref{Bquant}.

If $N=1$, the pair of energy shells is identified with the respective periodic orbits and the scar disks are merely the 2-D phase space interiors to each shell. Then the condition for $\Vxi(\x_{jj'})$ to lie on both periodic orbits is obtained automatically. Choosing quantized energies for both shells, one can sum over the factors $\exp[-\frac{\epsilon}{\hbar}(|t_j| + |t_{j'}|)]$ and then decrease the energy width $\epsilon$ arbitrarily 
to obtain the transition probability for the pair of individual eigenstates from the sharp peak of the transition density.

If $N>1$, the stationary centre $\x_{jj'}$ in the $(2N)$-D interior of both shells need not lie on a 2-D scar disk, 
but it can be so placed if there are sufficient parameters. The simplest case is that of the phase space reflections studied in {\bf I}.
Then the stationary point for a diagonal energy transition coincides with the reflection centre itself, so that one can merely place it 
anywhere on the desired scar disk. There is a shift of $\x_{jj'}$ from the reflection centre for a nondiagonal transition, 
but this can still be placed on a desired point 
of the scar disk by an appropriate choice within the $(2N)$-D space of reflection centres. In contrast, the single time parameter in a continuous flow generated by a fixed Hamiltonian is not sufficient to place the stationary point onto a given scar disk. 

If $N=2$, a single parameter together with the time specifies a 2-D manifold of stationary chord centres, which generically intersects a 2-D scar disk at isolated points, where full phase reinforcement is obtained. Defining the stability exponents of the quantized periodic orbit
with period $\tau_o$ as $\pm \lambda_o$, allows for the approximation of one of the amplitude factors in the transition density (4.10).
That is, for $\e^{k\lambda_o}>> 1$,
\be
|\det(\Id_\y+ \M_\y(t_j+k \tau_o))|^{-1/2} \approx \mu(t_j)^{-1/2} ~\e^{-k\lambda_o/2},
\label{stabpo}
\ee
where, expressing the stability matrix $\M_\y(t_j)$ for the shortest segment of the periodic orbit in the basis that diagonalizes the stability matrix for its full return, $\mu(t_j) \equiv \M_\y(t_j)_{11} $. Thus, summing this amplitude factor together with the
attenuation in time over repetitions of the exactly quantized orbit,
\be
\sum_{k=0}\frac{\exp\left[-\frac{\epsilon}{\hbar}(t_j + k\tau_o)\right]}{|\det(\Id_\y+ \M_\y(t_j+k \tau_o))|^{1/2}}
\approx \frac{\exp\left[-\frac{\epsilon}{\hbar}t_j\right] }{\mu(t_j)^{1/2}} 
~\frac{1}{1-\exp\left[-\lambda_o -\frac{\epsilon}{\hbar}\tau_o\right]}~.
\label{sumscar}
\ee
This does not become singular with $\epsilon \rightarrow 0$, as in the case where $N=1$, because of the presence of the stability exponent, 
but a marginally unstable quantized periodic orbit, with $\lambda_o$ close to unity, will strongly reinforce the energy transition density
if the stationary centre $\x_{jj'}$ lies exactly on its scar disk. To obtain the full amplitude factor for this point on the scar disk,
one adds
\be
\frac{\exp\left[-\frac{\epsilon}{\hbar}t_j \right]}{|\det(\Id_\y+ \M_\y(t_j))|^{1/2}}
-\frac{\exp\left[-\frac{\epsilon}{\hbar}t_j\right] }{\mu(t_j)^{1/2}} 
\ee 
to \eref{sumscar}, since the approximation \eref{stabpo} does not hold for the first trajectory segment.

The study of the decay of this amplitude reinforcement, as $\x_{jj'}$ is removed from a scar disk, is beyond the scope of the present paper.
There are interesting possibilities to be explored, such as the periodic orbits of arbitrarily high period that accumulate on homoclinic
and heteroclinic trajectories \cite{Douady}. Indeed, they differ in period by nearly $k\tau_o$ and in action by nearly the periodic orbit action,
i.e. their action difference is nearly quantized, even though they wonder off along the stable and unstable manifolds of the quantized orbit. 
Such contributions from orbits with different structure but similar action have been the source of fruitful advances in SC theory, starting from
the seminal paper by Sieber and Richter \cite{SiebRich}.

\section{Semiclassical scenario for energy transitions}

The stationary phase evaluation of individual compound orbit contributions should now be completed with a broader picture of the transition density. This relies on further generalizations of Poincar\'e sections to those presented in {\bf I}, following \cite{Report}.
Let us recall that the centre section of the $E$-shell is the intersection with its reflection, or that of another $E'$-shell, through a given point, here chosen as $\x_{jj'}$. The $j$-segment, contributing to the spectral Wigner function and hence to the above SC approximation, participates in the mapping of the section onto itself. Originally in {\bf I}, the reflected shells were identified with the driven shells, but this no longer holds 
in the present wider context. Yet one can still define the driven centre section in complete analogy, that is, 
the intersection of the driven $E'$-shell with its reflection through the same stationary point $\x_{jj'}$. Then it is the $j'$-segment, contributing to the driven spectral Wigner function, that participates in the $j'$-mapping of the driven centre section onto itself,
in which the lable $j'$ indicates the number of times the segment crosses the section. These centre sections are defined for any centre $\x$, but it is only at the stationary points $\x_{jj'}$ that both segments meet to form the closed $jj'$-compound orbit.  

The pairs of centre sections and their mappings may be considered as scaffolding, superseded by the alternative {\it evolved section}, resulting from the intersection of the $E$-shell with the driven the $E'$-shell. Clearly, the pair of stationary trajectory segments also have their tips on this section for all $j$ and $j'$ windings, so that one may consider the smooth extension to neighbouring trajectories, thus defining the $j$-mapping and the $j'$-mapping of the evolved section onto itself. It follows that all stationary contributions to the transition probability density are constructed on the fixed points of the product of a $j$-mapping with a $j'$-mapping of the evolved section onto itself. The common centre $\x_{jj'}$ of each of the related closed compound orbits depends on $j$ and $j'$. Both the $j$-segment on the $E$-shell and the $j'$-segment on the driven $E'$-shell are smooth solutions of their respective Hamilton equations, joined together by their endpoints, which lie on the evolved section.
It is stressed that no repetition of the compound orbit is included, unlike usual periodic orbit theory, since 
a single $j$-segment on the $E$-shell is followed by a single $j'$-segment on the $E'$-shell, no matter how many individual windings are indicated by $j$ and $j'$.  

The compound orbit determines a fixed point of a product of a $j$-mapping of (possibly) multiple traversals of the evolved section by a trajectory on the $E$-shell, followed by a $j'$-mapping by a trajectory on the $E'$-shell. Both these mappings will be chaotic, if so is the original Hamiltonian $H(\x)$, implying that the compound orbits will be just as isolated as the periodic orbits in the original $E$-shell. The families of evolved sections, along with their $jj'$-mappings, are identified by a single continuous parameter the time $\tau$, or $2N$ parameters such as either a translation $\Vxi$ or a reflection centre $\x$. Thus, on a smooth family of evolved sections are determined smooth families of $j$-mappings and $j'$-mappings, which give rise to smooth families of $jj'$-fixed points associated to compound orbits. It is then possible to follow the $jj'$- families of compound orbits numerically, like ordinary periodic orbit families \cite{Praguiar94}, converging by Newton's method for each small parameter increment, even through eventual bifurcations. 

The classical approximation constructed in {\bf I} for the centre section can be immediately extended to general evolutions. 
Indeed, the extra contribution of very short trajectories,
is not obtained by stationary phase, but it will be identified as the 00-term in the full SC expression for the transition density. This approximation holds where both spectral Wigner functions are adequately described by Dirac $\delta$-functions on their respective energy shells. Within this approximation, the integral (2.13) for the transition density is constrained to the evolved section and
\be
P_{EE'}(\tau,\epsilon)_{00} \approx \int d\x ~ \delta(H_-(\x) - E) ~ \delta(H_+(\x|\tau) - E').
\label{Ptrans2}
\ee
Thus, following {\bf I}, one obtains the generalized classical contribution as
\be
P_{EE'}(\tau,\epsilon)_{00} \approx \oint_{\x_s} d\x ~|\{H_-(\x),H_+(\x|\tau)\}|^{-1} ~,
\label{Ptrans4}
\ee
where the single Poisson bracket here matches the Hamiltonian with its evolved image, hence the change in notation. 
As discussed in {\bf I}, the integrand only becomes singular over the evolved section at caustics, but the result is finite if $N>1$.

The subgroup of translations $\Vxi(\tau) = \tau \Valpha$ generated by the driving Hamiltonian $\Lambda(\x)=\Valpha\wedge\x$ provides the simplest example of time evolutions, being that, as for Wigner functions, the quantum evolution of chord functions is purely classical. 
The chord function for a pure state $|\psi\rangle$ is
\be
\chi(\Vxi) = \frac{1}{(2\pi \hbar)^N} \langle\psi|{\hat T}_{2\xi} |\psi\rangle ~
\ee  
and, if the original Hamiltonian $H(\x)$ has a centre of symmetry, then chord functions for its eigenstates and their mixtures are real and
equal to the respective Wigner functions. 

Both these phase space representations of a symmetric eigenstate have their maximum at the origin. 
This peak is shared by the spectral Wigner functions and the diagonal term of the transition density (for no energy transition) 
is merely $|W_E(\x,\epsilon)|^2$. Semiclassically it becomes a supercaustic: Instead of being merely tangent at a single point, the yet to be evolved image of the energy shell coincides with it. Then, early in the evolution driven by $\Lambda(\x)=\Valpha\wedge\x$, it cuts the shell along a maximal section, such that any trajectory segment of the unperturbed shell reaching the section is the initial point of a returning segment along the translated shell that nearly reaches its initial point, that is, all points on the evolved section are nearly fixed points! There is then a minimal interval before the global participation of all the points on the evolved section condenses onto the individual SC contributions of selected trajectory segments. This is a very short transit, because the pair of segments are long and all that is needed is for the action between them to be of order $\hbar$. 

There are notable differences for a nondiagonal transition driven by a general evolution. A finite interval must elapse before the internal shell
touches the external shell at a local caustic, even in the case of a continuous group of translations. 
Unlike the external caustics discussed in {\bf I}, which  limit the parameters for which the driven $E'$-shell lies entirely outside the $E$-shell, this internal caustic marks the parameter for which they are no longer nested inside each other. Beyond this caustic parameter, the pair of segments forming the compound orbit quickly elongate to reach a growing section, which becomes nearly maximal as in the diagonal case. From then on, both cases are similar and the evolved section slowly shrinks again, until the external caustic is reached.
Even with the cutoff in time supplied by the finite energy widths, there will be many closed compound orbits that are fixed points of $jj'$-mappings of a large evolved section. However, as the section diminishes on the approach of an external caustic, fewer contributions of sufficiently small period cross the section and finally the transition density is dominated by the caustic contribution as in {\bf I}.

The SC approximation for the transition density in terms of  complex exponential oscillations becomes singular as either caustic is approached and it needs to be replaced by a uniform approximation, already discussed in {\bf I} in the case of phase space reflections. The reflection symmetry between the $E$-shell and its evolution simplifies the uniform approximation, which is expressed in terms of the Airy function \cite{Abramowitz}, but for the nondiagonal expression and generally for the chord function of nonsymmetric energy shells, one needs the complete fold-catastrophe expression including the derivative of the Airy function, as described by Berry \cite{Ber76} and implemented for chord functions in \cite{ZamOz10}. 
In any case, the approximation \cite{Report}
\be
S_j(\x_{jj'}) \approx \frac{2^{5/2}}{3}\frac{(E-H(\x_{jj'}))^{3/2}}{(\dot{\x}_{jj'}~ \mH_{jj'}~ \dot{\x}_{jj'})^{1/2}}
\ee
for the action of a contribution of a short trajectory segment to the spectral Wigner function if $\x$ lies close to the $E$-shell, with 
\be
\mH_{jj'} = \frac{\der^2}{\der\x^2} H(\x_{jj'})~,
\ee
leads to the dominant contribution to the Hessian matrix \eref{Hessian} as
\be
\B_{j} \approx  \sqrt\frac{2}{(E-H(\x_{jj'}))(\dot{\x}_{jj'}~ \mH_{jj'}~ \dot{\x}_{jj'})} ~~  \frac{\der H}{\der \x} \otimes \frac{\der H}{\der \x}~.
\ee
This is a large positive matrix near a caustic and $\B_{j'}$ will follow suit. Hence, the compound orbit contributions vanish on the caustic just as
the main classical contribution, as was shown in {\bf I}. Since these matrices are positive, the signature in (4.10) follows that of $-\B_j^2$, that is, $\sigma_{jj'}= -2N$. So it will remain, as the driving parameter changes continuously, until there is a zero eigenvalue in \eref{Hessall}, entailing a singularity in the SC approximation.

The fact that the spectral Wigner functions represent a mixed state for a classically narrow $\epsilon$-window of energies, effectively
a coarsegrained microcanonical ensemble, may seem a disadvantage. Taking $\epsilon \rightarrow 0$ to obtain a sharp transition rather than a density is problematic, as are resummation procedures to obtain a finite sum \cite{AgFish93,Report}. On the other hand, it should be borne in mind that an eventual future preparation in a real laboratory of an initial highly excited eigenstate may be quite difficult for a typical quantum system. The complete absence in a chaotic eigenstate of the very selection rules, which would allow an individual state to be picked within a dense spectrum, may be a serious obstacle.

\section*{Acknowledgments}
I thank Gabriel Lando for his help in preparing the figures.
Partial financial support from the 
National Institute for Science and Technology--Quantum Information
and CNPq (Brazilian agencies) is gratefully acknowledged.

\end{document}